\documentclass[conference]{IEEEtran}
\IEEEoverridecommandlockouts

\usepackage{amssymb}
\usepackage[cmex10]{amsmath}
\usepackage{stfloats}
\usepackage{graphicx}
\usepackage{subfigure}
\usepackage{tabularx}
\usepackage{epsfig,epsf,color,balance,cite}
\usepackage{verbatim}
\usepackage{url}
\usepackage{bm}
\usepackage{booktabs}

\usepackage{algorithm}
\usepackage{algorithmic}

\usepackage{color}
\definecolor{myc1}{rgb}{0,0,0}

\begin{document}

\title{Fair Resource Allocation for Probabilistic Semantic Communication in IIoT}

\author{
\IEEEauthorblockN{Siyun Liang\IEEEauthorrefmark{1},
                  Zhouxiang Zhao\IEEEauthorrefmark{1},
                  Chen Zhu\IEEEauthorrefmark{2},
                  Zhaohui Yang\IEEEauthorrefmark{1}\IEEEauthorrefmark{3},
                  Yinchao Yang\IEEEauthorrefmark{4},
                 \\ Mohammad Shikh-Bahaei\IEEEauthorrefmark{4},
                  and Zhaoyang Zhang\IEEEauthorrefmark{1}\IEEEauthorrefmark{3}
                 }
	\IEEEauthorblockA{
			$\IEEEauthorrefmark{1}$College of Information Science and Electronic Engineering, Zhejiang University, Hangzhou, China\\
                $\IEEEauthorrefmark{2}$Polytechnic Institute, Zhejiang University, Hangzhou, Zhejiang, 310015, China\\
			$\IEEEauthorrefmark{3}$Zhejiang Provincial Key Laboratory of Info. Proc., Commun. \& Netw. (IPCAN), Hangzhou, China\\ 
                $\IEEEauthorrefmark{4}$Department of Engineering, King's College London, London, UK\\
			E-mails: 
   \{siyunliang, zhouxiangzhao, zhuc, yang\_zhaohui, ning\_ming\}@zju.edu.cn, 
   \{yinchao.yang, m.sbahaei\}@kcl.ac.uk
		}
}

\maketitle

\begin{abstract}
In this paper, the problem of minimum rate maximization for probabilistic semantic communication (PSCom) in industrial Internet of Things (IIoT) is investigated. In the considered model, users employ semantic information extraction techniques to compress the original data before sending it to the base station (BS). During this semantic compression process, knowledge graphs are employed to represent the semantic information, and the probability graph sharing between users and the BS is utilized to further compress the knowledge graph. The semantic compression process can significantly reduce the transmitted data size, but it inevitably introduces additional computation overhead. Considering the power budget of the user, we formulate a joint communication and computation optimization problem aiming to maximize the minimum equivalent rate among all users while meeting total power and semantic compression ratio constraints.
To address this problem, two algorithms with different computational complexities are proposed to obtain sub-optimal solutions. One algorithm is based on a prorate distribution of transmission power, while the other traverses the combinations of semantic compression ratios among all users. In both algorithms, bisection is employed in order to achieve the greatest minimum equivalent rate.
The simulation results validate the effectiveness of the proposed algorithms.
\end{abstract}

\IEEEpeerreviewmaketitle

\section{Introduction}
The rapid evolution of wireless communication technology has led to the emergence of the industrial Internet of Things (IIoT) as a key focus in the current era. 
According to \cite{2618724}, a significant portion of the IIoT is occupied by low-cost wireless access and low power wide area network, which requires abundant communication capability while saving power throughout the communication process. In light of these considerations, semantic communication has been introduced to the communication in IIoT \cite{10550151}. This approach contributes to energy conservation in wireless networks in IIoT and opens up new possibilities in a semantic view.

Research in semantic communication appears a departure from existing communication models \cite{liang2023generative}. However, the paradigm of semantic communication still lacks a clear classification and remains to be evolved.

The nature of semantic communication suits great with wireless networks. In \cite{8461983}, deep learning techniques were introduced to joint source-channel coding of text. This combination laid the foundation of semantic communication systems for text transmission assisted by deep learning. Futhermore, the work in \cite{8461983} offered a novel way to encode transmitting data with less distortion over the whole communication system. With this forward research, an iterative semantic coding method of text was introduced in  \cite{9834044}. And the work in \cite{10024766} introduces the application of edge learning (EL) in distributed goal-oriented semantic communication. As mentioned before, the main purpose of semantic coding is to capture the underlying meaning of messages, thus redundant data that are irrelevant in the context of transmission, are spared from transmission, as shown in \cite{10233741}. To better show the semantic relations between the entities hidden in the context, semantic triples and knowledge graphs have been employed. Liu et al. investigated a task-oriented semantic communication approach based on semantic triples \cite{10118916}. The work in \cite{10118916} mainly focus on developing an approach which is able to transmit the needed information according to the requirements of a specific task. Meanwhile, the authors in \cite{9838470} adopted knowledge graphs as an approach and proposed a cognitive semantic communication framework. This work proposed a simple, general, and interpretable semantic alignment algorithm to detect the semantic information hidden the plain message. Furthermore, the work in \cite{e26050394} proposed an optimization method in an multi-user probabilistic semantic communication (PSCom) framework with a base station (BS) equipped with multiple antennas. On accounted of the semantic detecting and computation in semantic communication, the computation load cannot be neglected comparing to the transmitting load. Thus, resource allocation and performance optimization are decisive in a semantic communication system. The problem was addressed in \cite{9832831}, which employed deep reinforcement learning in a semantic communication framework. It is worth-noting that the aforementioned works \cite{8461983,9834044,10118916,9838470,9832831} did not take computation load and resource allocation into consideration, while \cite{e26050394} and \cite{9707474} had proven that these issues are important for energy-constrained wireless networks.

The primary highlight of this research is the development of a multi-user PSCom framework applicable to IIoT. This model considers both transmission and computation consumption in an integrated manner. The essential contributions of our work can be summarized as follows:
\begin{itemize}
\item We analyze a PSCom network where multiple users implement semantic information extraction techniques. These techniques are used to compress the users' original high-volume data, which is then transmitted to a BS. In our framework, the users' large-sized data is represented by comprehensive knowledge graphs. They are then compressed based on the shared probability graph existing between the users and the BS.
\item We formulate an optimization problem with the aim of maximizing the minimal equivalent rate across all users, with considering the total power constraint. This problem considers the balance between transmission proficiency and the intricacy of computation. With this optimization target, we are able to guarantee the fairness of power allocation between users, without leaving one with comparatively worse equivalent rate.
\item To solve this problem, two algorithms with different computational complexities are presented. Numerical results show the effectiveness of the proposed algorithms.
\end{itemize}

\section{System Model and Problem Formulation}
\begin{figure}[t]
\centering
\includegraphics[width=0.9\linewidth]{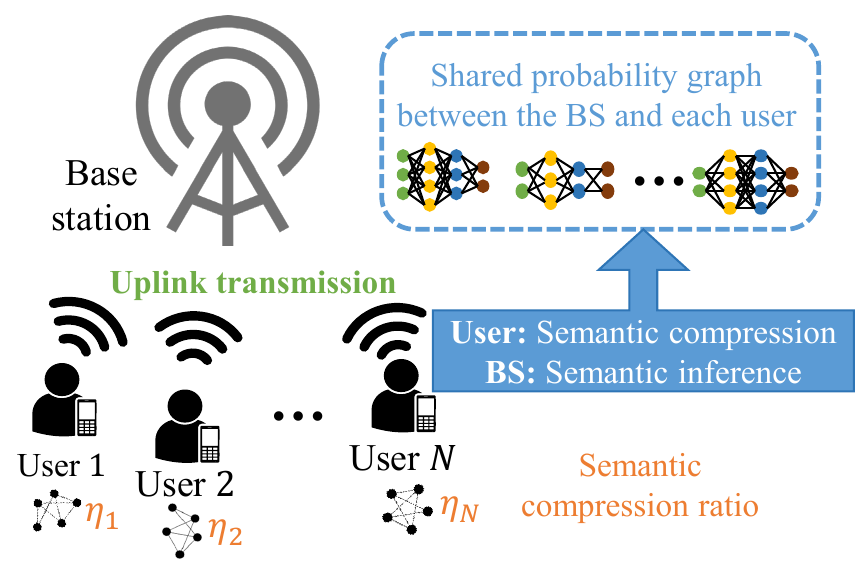}
\caption{The considered PSCom network.}
\label{fig:1}\vspace{-1em}
\end{figure}

Consider an uplink wireless PSCom network with one BS and $N$ users, as shown in Fig.~\ref{fig:1}. The set of users is denoted by $\mathcal{N}$. Each user, denoted by $n$, has a large-sized data $\mathcal{X}_n$ to be transmitted. Due to limited wireless resource, the users need to extract the small-sized semantic information $\mathcal{Y}_n$ from the original data $\mathcal{X}_n$. In the considered model, users first extract the semantic information based on their individual local probability graphs, and then transmit the extracted information to the BS.

\subsection{Semantic Communication Model}
\begin{figure}[t]
\centering
\includegraphics[width=\linewidth]{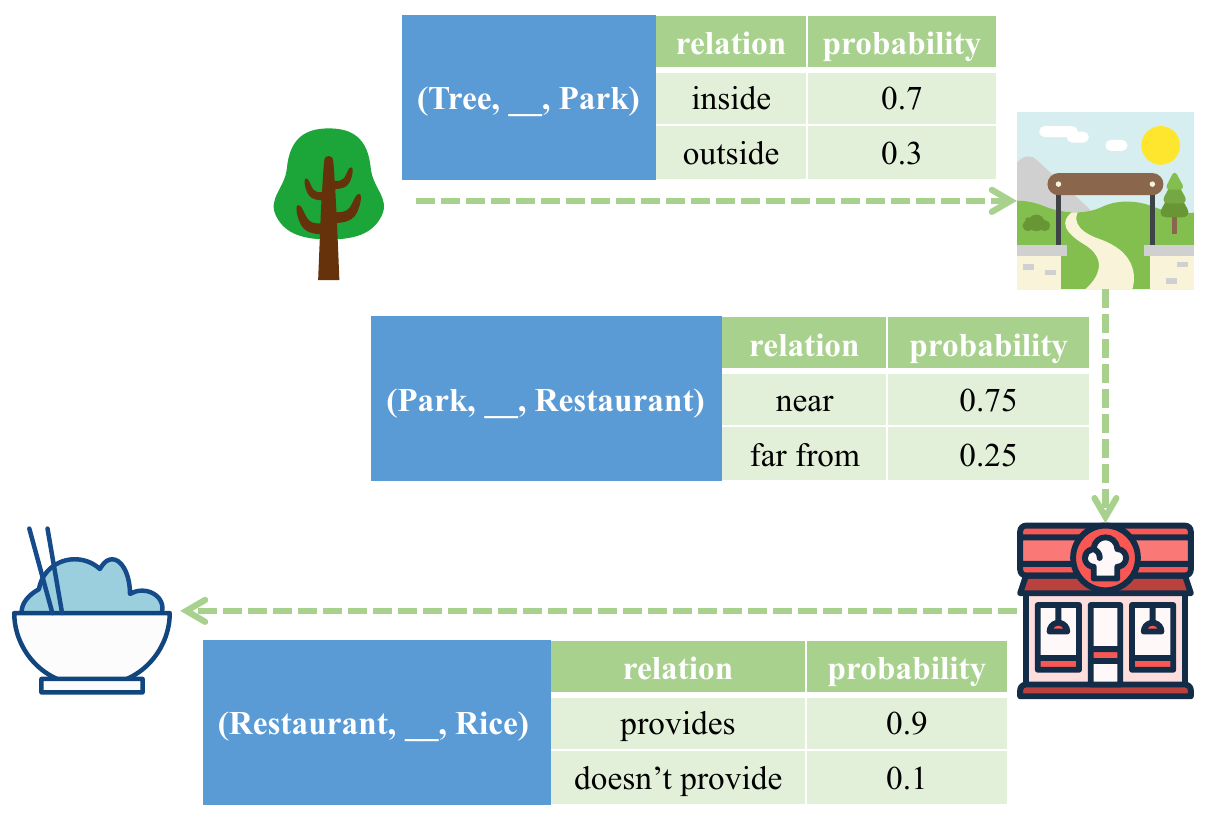}
\caption{An illustration of a probability graph in the considered PSCom system.}
\label{fig:2}
\end{figure}

\begin{figure}[t]
\centering
\includegraphics[width=\linewidth]{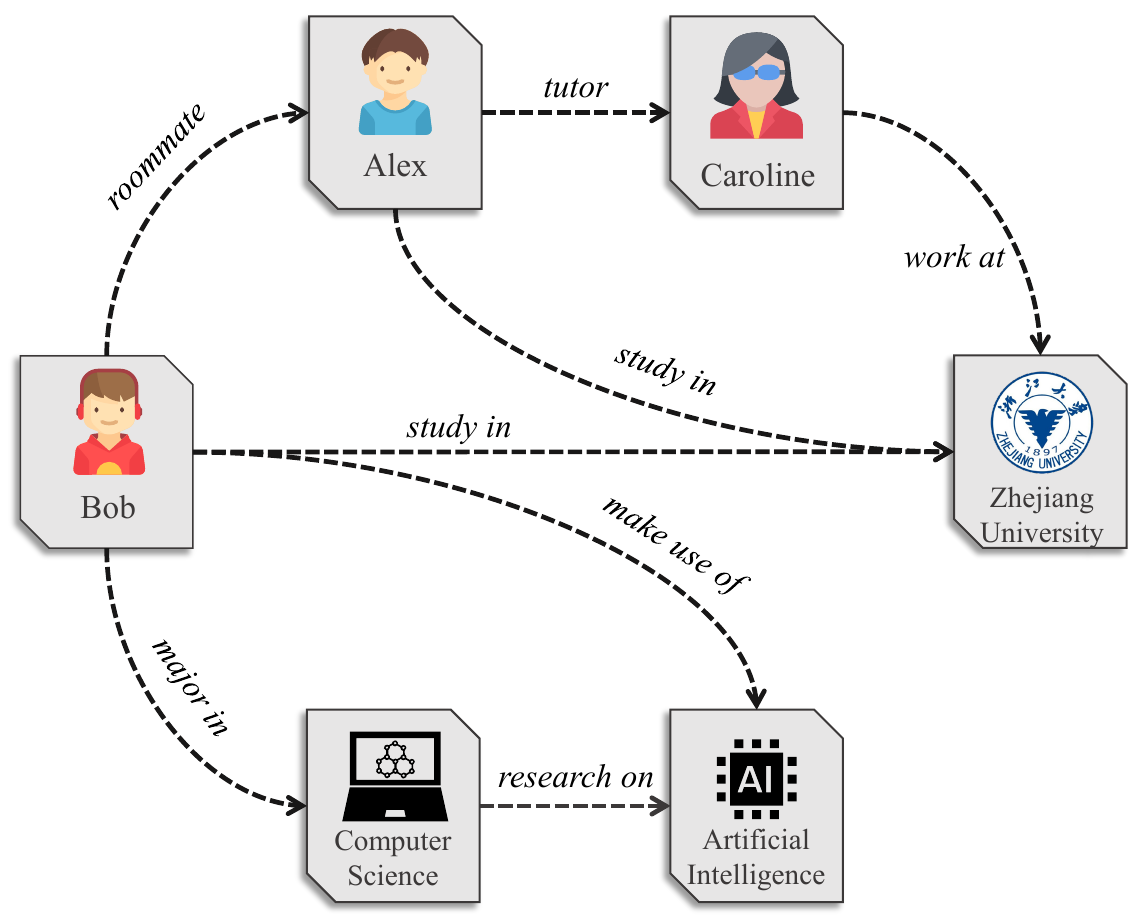}
\caption{An illustration of a knowledge graph.}
\label{fig:3}\vspace{-1em}
\end{figure}

In this paper, we use probability graphs as the connecting knowledge base between the semantic transmitter (user side) and the semantic receiver (BS). An illustrative example of a probability graph is depicted in Fig.~\ref{fig:2}. A probability graph merges data from multiple knowledge graphs. Comparing to conventional knowledge graphs, probability graphs connect the entities with relational probability. A standard knowledge graph consists of a multitude of triples, each of which can be represented by $\varepsilon = (h, r, t)$, where $h$ represents the head entity, $t$ stands for the tail entity, and $r$ is the relation between $h$ and $t$. An example of a probability graph is shown in Fig.~\ref{fig:3}. Generally, in a knowledge graph, the relations are stable. Meanwhile, in a probability graph, each relation corresponds to a relational probability, indicating the probability of that specific relation occurring under the conditions of a fixed head entity and tail entity.

Within the proposed framework, we assume that every user possesses local knowledge graphs for transmission. These knowledge graphs are created from comprehensive textual data (alongside possible picture/audio/video data). To process the data, named entity recognition (NER) as well as relation extraction (RE) are used, leading to high-density information. By leveraging the common probability graph between a user and the BS, the knowledge graphs can be further compressed.

The probability graph widens the scopes of relations by statistically accounting for various relational occurrences tied to the same head and tail entities across different instances of knowledge graphs. The statistical information from the probability graph can be used to create a multidimensional conditional probability matrix. Such a matrix embodies the probability that a specific triple is deemed valid under the certain triples are valid. This availability allows us to decrease certain relations in the knowledge graph during transmission, leading to data compression.

However, it's important to note that obtaining a smaller data size requires a reduced semantic compression ratio, which in turn needs larger-dimensional conditional probabilities. This reduction in semantic compression ratio implies an increase in computational load, which forms a trade-off between transmission and computation load for the mentioned PSCom network. The specific implementation details of the probability graph can be found in \cite{ZHAO2024107055}.

In the considered PSCom network, each user possesses a unique local probability graph, which captures statistical information about their past data. Each user $n$, carries out semantic information extraction on an individual basis, compressing the original large-scale data $\mathcal{X}_n$ according to its stored probability graph. This is done at a semantic compression ratio indicated by $\eta_n$. Subsequently, the data denoted by $\mathcal{Y}_n$, is sent to the BS with a transmission power of $p_n^\mathrm{t}$.

Simultaneously, the BS maintains identical probability graphs that correspond to all $N$ users. Upon receiving the semantic data from the $n$-th user, the BS performs a semantic inference to restore the compressed semantic information, leveraging the shared probability graph of $n$-th user. 

\subsection{Transmission Model}
We consider the uplink transmission from all users to the BS and assume the channel to be an AWGN channel. According to Shannon's formulation, the capacity of the channel between the BS and the $n$-th user can be expressed as:

\begin{equation}
    C_n=B\log_2 (1+\frac{p_n^\mathrm{t} h_n}{\sigma^2}),
\end{equation}
where $p^\mathrm{t}_n$ is the transmission power for the $n$-th user, $h_n$ is the channel gain, and $\sigma^2$ is the noise power among the channel.

In the context of PSCom network, the original large-scale data $\mathcal X_n$ is compressed into a smaller scale data $\mathcal Y_n$ with a semantic compression rate before transmission. Using a function $\mathrm{size}(\cdot)$ to measure the data size in terms of bits, the semantic compression ratio for the $n$-th user is defined as follows:
\begin{equation}\label{cr}
    \eta_n=\frac{\mathrm{size}(\mathcal{Y}_n)}{\mathrm{size}(\mathcal{X}_n)}.
\end{equation}

Consequently, we are able to obtain the equivalent rate for the $n$-th user:
\begin{equation}\label{eq:eqr}
    R_n=\frac{1}{\eta_n}C_n.
\end{equation}

In comparison to traditional channel capacity, \eqref{eq:eqr} is multiplied by a factor $\frac{1}{\eta_n}$, because the data transmitted in the channel is pre-compressed at the compression rate $\eta_n$. As such, the size of data after decoding at the receiver end is anticipated to be $\frac{1}{\eta_n}$ times the size before decoding.

\subsection{Computation Model}
In this PSCom network, we perform semantic information extraction drawing upon the user's local probability graph to compress the original data of the $n$-th user, $\mathcal{X}_n$, into a smaller-sized data set, denoted $\mathcal{Y}_n$. It's important to note that the compression process itself calls for computational resources. Furthermore, the better the compression result (i.e., the lower the compression rate $\eta_n$), the larger the computational power required.

According to equation (20) in \cite{ZHAO2024107055}, the computational load for the PSCom network under consideration can be expressed in the following way:
\begin{equation}\label{cl}
    g\left(\eta\right)=\left\{\begin{array}{l}
        A_1\eta +B_1, D_1\leq \eta \leq 1, \\
        A_2\eta +B_2, D_2\leq \eta < D_1, \\
        \vdots \\
        A_S\eta +B_S, D_S\leq \eta < D_{S-1},
    \end{array}\right..
\end{equation}

Based on \eqref{cl}, this computation load function is made up of $S$ segments. Each segment is linear. Here, $A_s$ indicates the slope for the $s$-th segment, while $B_s$ denotes the constant in the linear function. The function reflects the nature of power consumption in that the lower the compression rate $\eta_n$ is, the more power is used. Hence, the slope $A_s$ must be negative. The pair $(D_s, D_{s-1})$ describes the value range of the $s$-th segment. According to \cite{ZHAO2024107055}, these segments arise from successive levels of conditional probability functions.

\begin{figure}[t]
\centering
\includegraphics[width=0.8\linewidth]{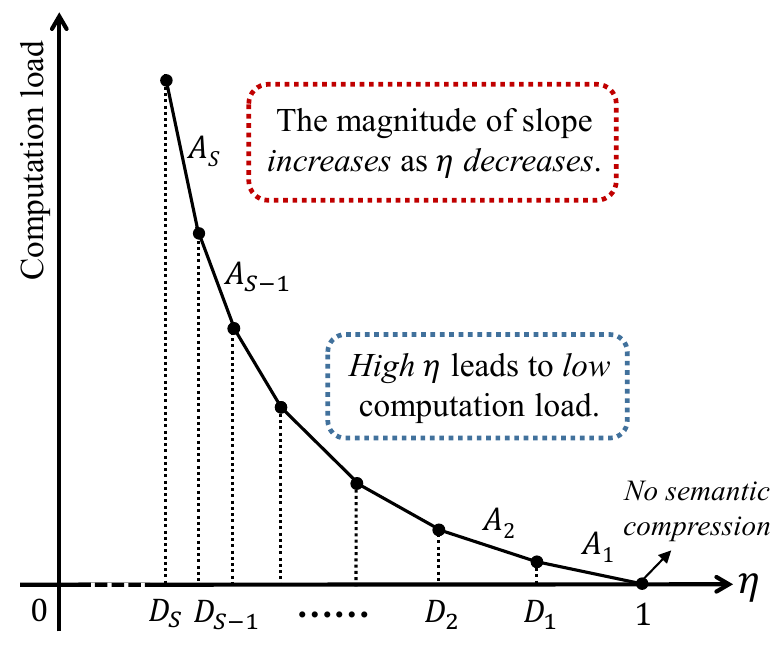}
\caption{Illustration of computation load vs. semantic compression ratio $\eta$.}
\label{fg:cl}\vspace{-1em}
\end{figure}

Indeed, as shown in equation \eqref{cl}, the computation load, represented by $g(\eta)$, showcases a segmented structure spanning $S$ levels, with the slope magnitude exhibiting discrete decrements, as visualized in Fig.~\ref{fg:cl}. Consequently, when operating at a high compression ratio, we use only low-dimensional conditional probabilities, which leads to a reduction in computational demands. Conversely, as the compression ratio falls, there's an increased demand for higher-dimensional information, which imposes a more intensive computational load. Significantly, each inflection point in the segmented function $g(\eta)$ marks the deployment of probabilistic information characterized by more comprehensive data for the extraction of semantic information.

With the expression of computation load function $g(\eta)$ in \eqref{cl}, we can write the compression power for the $n$-th user as:
\begin{equation}\label{cp}
    p_n^c = g_n(\eta_n)p_0,
\end{equation}
where $p_0$ stands for a positive constant denoting the computation power coefficient.

In this paper, our aim is to balance the power resources between the computational processes happening at the user end and the transmission taking place between the user and the BS. The objective is to ensure that the lowest equivalent rate for all users is maintained, while operating under a specified total power cap.

\subsection{Problem Formulation}
Given the considered system model, our aim is to maximize the minimum equivalent rate among all users. The equivalent rate for the $n$-th user can be represented as
\begin{equation}\label{eq:r}
    R_n=\frac{B}{\eta_n}\log_2\left(1+\frac{p_n^\textrm{t} h_n}{\sigma^2}\right).
\end{equation}

Denote the minimum equivalent rate among all users as $\tau$, i.e., $\tau=\min\limits_{n} R_n, \forall n \in \mathcal{N}$, the problem can be formulated as
\begin{subequations}
    \begin{align}
        \max_{\bm{\eta}, \mathbf{p^\mathrm{t}}} \quad & \tau, \label{eq:tau} \\
        \mathrm{s.t.} \quad & \sum_{n=1}^N p_n^\mathrm{t}+p_n^\mathrm{c}\leq P^\mathrm{max}, \label{eq:c1}\\
        & p_n^\mathrm{t}\geq 0,\forall n\in\mathcal{N}, \label{eq:c2}\\
        & 0<\eta_n\leq 1,\forall n\in\mathcal{N}, \label{eq:c3}
    \end{align}
\end{subequations}
where $P^\mathrm{max}$ represents the overall power constraint for all users.

Constraint \eqref{eq:c1} introduces an upper limit to the collective power used for both transmission and computation. Constraint \eqref{eq:c2} guarantees non-negative transmit power for every user. Lastly, constraint \eqref{eq:c3} sets boundaries for the semantic compression ratio for each user.

\section{Algorithm Design}

Due to the complex coupling interaction between $\eta_n$ and $p_n^\mathrm{t}$, deriving an optimal solution for this problem presents significant challenges. Consequently, we proposed two algorithms that allow us to achieve a sub-optimal solution with decreased complexity.

This approach enables us to deconstruct the main problem into two smaller, more manageable sub-problems. These can then be addressed independently, as follows:

\subsection{Method-1}

For the optimal solution of max-min problem \eqref{eq:tau}, we always have that the equivalent rate of each user is the same.  With the optimal $\tau$ in \eqref{eq:tau}, we can have:
\begin{equation}
    \tau = \frac{B}{\eta_n}\log_2(1+\frac{p_n^\mathrm{t} h_n}{\sigma^2}).
\end{equation}

The compression rate $\eta_n$ can be written as:

\begin{equation}\label{eq:rn}
    \eta_n=\frac{B}{\tau}\log_2(1+\frac{p_n^\mathrm{t} h_n}{\sigma^2}).
\end{equation}

When $\tau$ is optimized, a predetermined $p_n^\mathrm{t}$ will result in a specific $\eta_n$. Assuming this, we can conduct a bisection with an initial transmission power $p_n^\mathrm{t}$.

To start, we introduce a method to set an initial transmission power $p_n^\mathrm{t}$. We represent $p_n^\mathrm{t}$ in terms of the relevant channel gain $h_n$ and a power proportion $\beta$:
\begin{equation}\label{eq:pb}
    p_n^\mathrm{t}=\frac{\beta}{h_n}.
\end{equation}

With the constraint \eqref{eq:c1} and $p_n^\mathrm{c} \geq 0$, it is clear that $\beta$ must be subject to the constraint:

\begin{equation}\label{eq:range}
     0 \leq \beta \leq \frac{P^\mathrm{max}}{\sum^\mathcal{N}_{n=1} 1/h_n}.
\end{equation}

In this paper, we equidistantly sample from the range \eqref{eq:range} for $m$ different values to be the power proportion $\beta$ and select the maximum result as the final one.

On the whole, \eqref{eq:pb} reveals how the allocation of power to a user is dependent on the respective channel condition. If the channel condition is deteriorating, additional power will be assigned to the transmission.

According to \eqref{eq:rn} and \eqref{eq:pb}, we can obtain $\eta_n$ for every $\tau$ and initialized $p_n^\mathrm{t}$. With the correspondence between $\tau$ and $\eta_n$, we can rewrite constraint \label{c1} as follows:
\begin{equation}\label{eq:sum}
    \sum_{n=1}^{\mathcal{N}} p_n^\mathrm{c}(\tau)+p_n^\mathrm{t}\leq P^\mathrm{max}.
\end{equation}

Therefore, the final $\tau$ that we derive should adhere to equation \eqref{eq:sum}. With this limitation in place, we can execute a bisection. It starts with a given section, defined by the lower bound $\tau_{\mathrm{\min}}$ and upper bound $\tau_{\mathrm{\max}}$ respectively. Then it keeps shrinking the section by lowering $\tau_{\mathrm{\max}}$ or magnifying $\tau_{\mathrm{\min}}$, until the gap between $\tau_{\mathrm{\min}}$ and $\tau_{\mathrm{\max}}$ is smaller than a given threshold $\epsilon$. Details are outlined in Algorithm 1.
\begin{algorithm}
\caption{Bisection for optimizing $\tau$ with fixed transmission power}
    \begin{algorithmic}
        \FORALL{power proportion $\beta$}
        \STATE Initialize the searching range of $\tau$ with a maximum value $\tau_{\mathrm{\max}}$ and minimum value $\tau_{\mathrm{\min}}$
        \WHILE{$\tau_{\mathrm{\max}}$-$\tau_{\mathrm{\min}}<\epsilon$}
            \STATE$\tau=\frac{\tau_{\mathrm{\max}}+\tau_{\mathrm{\min}}}{2}$
            \STATE Calculate $\bm{\eta}$ with  \eqref{eq:rn}.
            \STATE Calculate $p_{sum}=\sum_{n=1}^{\mathcal{N}} p_n^\mathrm{t}+p_n^\mathrm{c}$
            \IF{$p_{sum}>p_{\mathrm{\max}}$}
            \STATE $\tau_{\mathrm{\max}}=\tau$
            \ELSIF{$p_{sum}<p_{\mathrm{\max}}$}
            \STATE $\tau_{\mathrm{\min}}=\tau$
            \ELSE
            \STATE Break the loop
            \ENDIF
        \ENDWHILE
        \IF{The current obtained $\tau$ is larger than that obtained before}
            \STATE Update $\tau$
        \ENDIF
        \ENDFOR
        \STATE Find the maximum $\tau$.
    \end{algorithmic}
\end{algorithm}
\subsection{Method-2}
To simplify the issue, we employ the start and end points of each segment of the computation load function, $g(\eta)$, as the prospect values for $\eta_n$. We then implement a bisection for each vector set, $\bm{\eta}$.
\begin{equation}
    \bm{\eta}_n= l, l \in \{D_1, D_2 \cdots, D_{S}\}, n \in \mathcal{N}.
\end{equation}

It is worth noting that with every determined vector set $\bm{\eta}$, the computation load $\sum_{n=1}^{\mathcal{N}} p_n^\mathrm{c}$ for all users is fixed. 

Analogous to `Method-1', we can represent transmission power $p_n^\mathrm{t}$ employing an optimal $\tau$ and a fixed $\eta_n$:
\begin{equation}\label{eq:pt}
    p_n^\mathrm{t}=(2^{\frac{\tau \eta_n}{B}}-1)\frac{\sigma^2}{h_n}.
\end{equation}

It should be noted that the transmission power can be viewed as a function of $\tau$. Identical to \eqref{eq:sum}, the constraint for the $\tau$ that is obtained in `Method-2' is:
\begin{equation}\label{eq:sum2}
    \sum_{n=1}^{\mathcal{N}} p_n^\mathrm{c}+p_n^\mathrm{t}(\tau)\leq P^\mathrm{max}.
\end{equation}

Base on \eqref{eq:sum2} we can perform a bisection to find the maximum $\tau$. Similar to algorithm 1, the overall algorithm is summarized in Algorithm 2.
\begin{algorithm}
\caption{Bisection for optimizing $\tau$ with fixed computation load}
    \begin{algorithmic}
        \FORALL{computation load vector $\bm{\eta}$}
        \STATE Initialize the searching range of $\tau$ with a maximum value $\tau_{\mathrm{\max}}$ and minimum value $\tau_{\mathrm{\min}}$
        \WHILE{$\tau_{\mathrm{\max}}$-$\tau_{\mathrm{\min}}<\epsilon$}
            \STATE$\tau=\frac{\tau_{\mathrm{\max}}
            +\tau_{\mathrm{\min}}}{2}$
            \STATE Calculate $p^\mathbf{t}$ with  \eqref{eq:pt}.
            \STATE Calculate Calculate $p_{sum}=\sum_{n=1}^{\mathcal{N}} p_n^\mathrm{t}+p_n^\mathrm{c}$
            \IF{$p_{sum}>p_{\mathrm{\max}}$}
            \STATE $\tau_{\mathrm{\max}}=\tau$
            \ELSIF{$p_{sum}<p_{\mathrm{\max}}$}
            \STATE $\tau_{\mathrm{\min}}=\tau$
            \ELSE
            \STATE Break the loop
            \ENDIF
        \ENDWHILE
        \IF{The current obtained $\tau$ is larger than that obtained before}
            \STATE Update $\tau$
        \ENDIF
        \ENDFOR
        \STATE Find the maximum $\tau$.
    \end{algorithmic}
\end{algorithm}
\section{Simulation Results and Analysis}
This section presents a comparison between the `Method-1' and `Method-2' schemes, as well as two additional schemes designed to enhance the observation of results.

\begin{itemize}
    \item \textbf{Equal power allocation}: In this scheme, the total maximum power is distributed equally among all users, irrespective of their individual channel gain. Further, each user's equivalent rate is optimized using the gradient descent method.
    \item \textbf{Non-semantic}: In this scheme, no semantic compression will be executed. The total power is allocated to users in a manner inversely proportional to the channel gain.
\end{itemize}

The parameters of simulation, if not specified, are demonstrated in Table \ref{t:1}.

\begin{table}[ht]
\centering
\caption{Main System Parameters}
\label{t:1}
\begin{tabular}{ccc}
\hline
\textbf{Parameter}                  & \textbf{Symbol} & \textbf{Value} \\ \hline
Total maximum power & $P^\mathrm{max}$       & 6 W           \\
Channel noise power                 & $\sigma^2$      & -90 dBm           \\
Bandwidth                    & $B$             & 10 MHz             \\
Initial searching range upper limit & $\tau_{\max}$   & $10^{-3}$        \\
Initial searching range lower limit & $\tau_{\min}$   & $10^{10}$         \\
Stop threshold                      & $\epsilon$      & $10^{-4} $      \\
Computation power coefficient       & $p_0$          & $10^{-3}$              \\
Numbers of power proportion picked      & $m$          & 500              \\ \hline
\end{tabular}
\end{table}

\begin{figure}[t]
    \centering
    \includegraphics[width=\linewidth]{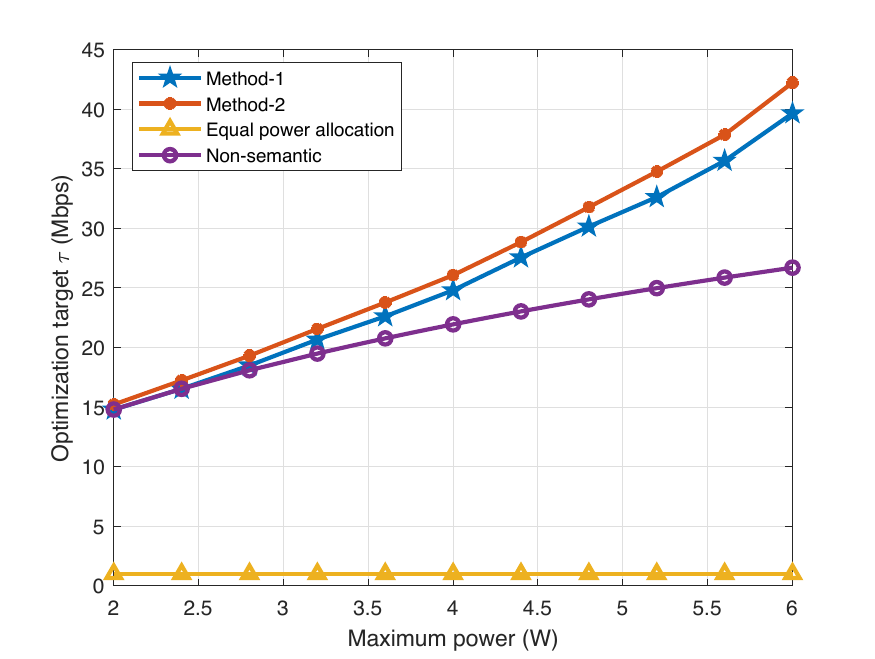}
    \caption{Maximum power vs. Optimization target $\tau$}
    \vspace{-1em}
    \label{fig:pmax}
\end{figure}

In Fig.~\ref{fig:pmax}, we compare the results of the four aforementioned schemes in relation to varied total maximum power. In all schemes, the optimization target—that is, the minimum equivalent rate among all—increases with an uptick in maximum power. However, when compared to the other schemes, the `Non-semantic' scheme lags significantly and its growth rate is considerably smaller. 

Both schemes `Method-1' and `Method-2' outperform the other two, with scheme `Method-2' demonstrating a slight edge over scheme `Method-1'. This lead, however, comes at the expense of increased computational complexity. As the number of users increases, there will be more instances where different users require different compression rates, which makes scheme `Method-2' attempt more situations while having a minimal impact on scheme `Method-1'.

\begin{figure}[t]
    \centering
    \includegraphics[width=\linewidth]{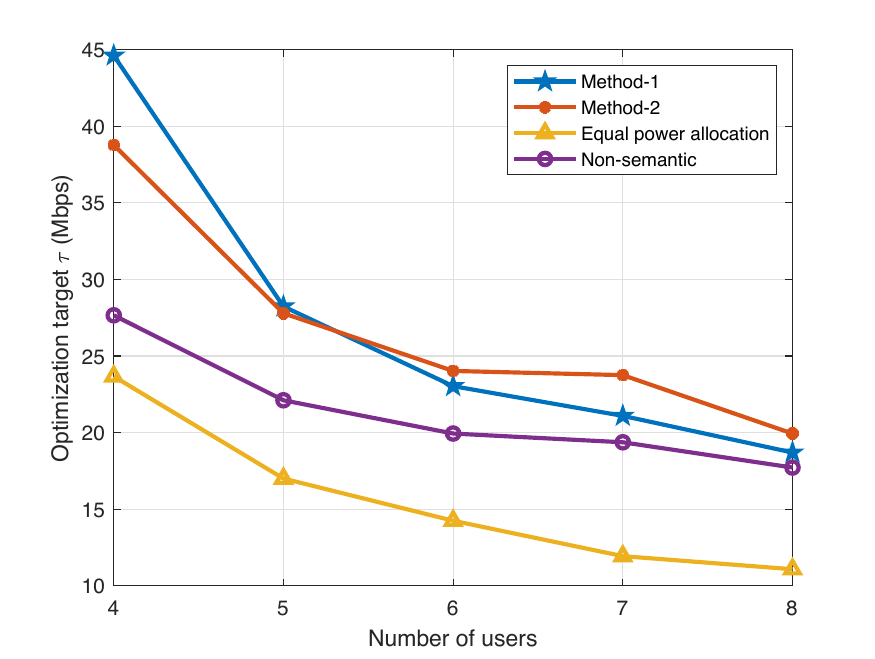}
    \caption{Number of users vs. Optimization target $\tau$}
    \vspace{-1em}
    \label{fig:user}
\end{figure}
In Fig.~\ref{fig:user}, we modify the number of users with different channel gains when the total power $P^\mathrm{max}$ fixed at $6 W$, and monitor the performance of the four schemes mentioned earlier. 

It is evident from the various schemes that an increase in the number of users results in a deterioration of the optimization target due to the limited sum of power. Generally, schemes `Method-1' and `Method-2' maintain their performance across all the schemes. As the number of users reaches 7, the discrepancies between schemes except `Equation power allocation' become negligible. Insufficient power impairs the efficacy of semantic compression, thus prompting both schemes `Method-1' and `Method-2' to allot the majority of their resources to transmission, which renders their performance indistinguishable from that of scheme `Non-semantic'. 

\begin{figure}[t]
    \centering
    \includegraphics[width=\linewidth]{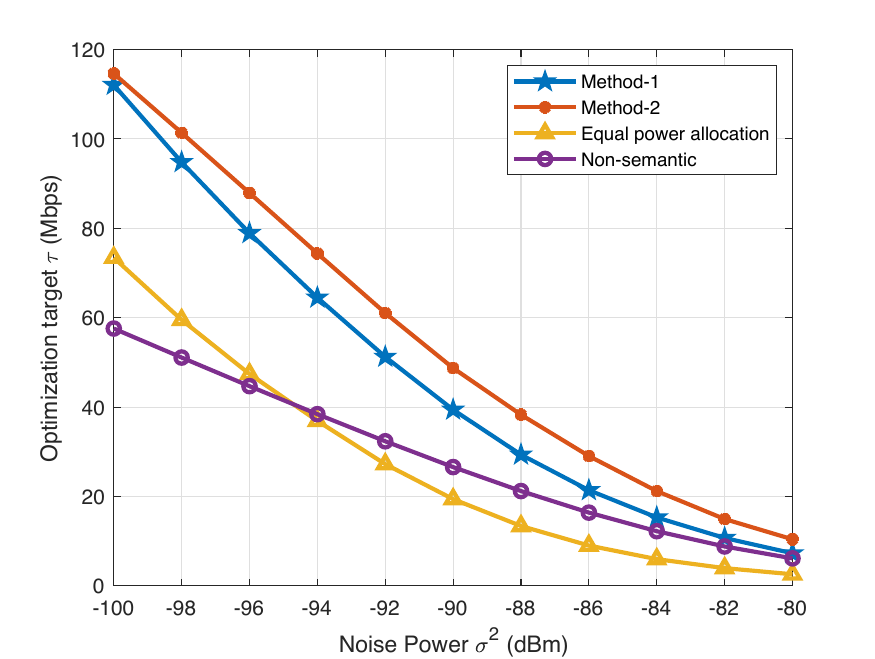}
    \caption{Noise Power vs. Optimization target $\tau$}
        \vspace{-1em}
    \label{fig:np}
\end{figure}

In Fig.~\ref{fig:np}, we examine the performances of the four schemes in the context of varying channel noise power. It is evident that as the noise power increases, a trend emerges showing a decrease in the minimum equivalent rate. Notably, scheme `Non-semantic' shows a distinctive slower decline compared to the other schemes. This can be explained intuitively by the fact that in scheme `Non-semantic', all power is allocated to transmission. Consequently, it has a better noise immunity.

\section{Conclusion}
In this paper, we analyze a PSCom network between multi-user and a BS, using semantic triples and knowledge graphs. Leveraging semantic communication, we are able to extract the underlying meaning in messages and simplify the communication content by introducing semantic compression, thus leading to lower the energy consumption. However, it leads to a complex communication network with local computation and remote transmission, which calls for a performance optimization when adopting the network. In this work, we seek to identify the optimal solution for maximizing the minimum equivalent rate among all users. To this end, we propose two algorithms that search for sub-optimal solutions to the problem. The effectiveness of the proposed algorithms is demonstrated through numeric results.

\bibliographystyle{IEEEtran}
\bibliography{main}
\end{document}